\documentclass[prl,twocolumn,showpacs]{revtex4}
\usepackage{graphicx,epsf}
\usepackage{bm,bbm}      

\newcommand{\BEDT}{$\alpha-$(BEDT-TTF)$_2$I$_3$}

\newcommand{\Pib}{\mbox{\boldmath $\Pi $}}

\newcommand{\bq}{{\bf q}}

\newcommand{\br}{{\bf r}}
\newcommand{\be}{{\bf e}}

\newcommand{\bv}{{\bf v}}
\newcommand{\bw}{{\bf w}}

\newcommand{\bA}{{\bf A}}
\newcommand{\bB}{{\bf B}}
\newcommand{\bE}{{\bf E}}

\newcommand{\wtilde}{\tilde{w}}

\newcommand{\beq}{\begin{equation}}
\newcommand{\beqn}{\begin{eqnarray}}
\newcommand{\eeq}{\end{equation}}
\newcommand{\eeqn}{\end{eqnarray}}
\newcommand{\nn}{\nonumber}

\begin{document}

\def\tende#1{\,\vtop{\ialign{##\crcr\rightarrowfill\crcr
\noalign{\kern-1pt\nointerlineskip}
\hskip3.pt${\scriptstyle #1}$\hskip3.pt\crcr}}\,}

\title{Electric-Field-Induced Lifting of the Valley Degeneracy in \BEDT~Dirac-Like Landau Levels}
\author{M.O. Goerbig, J.-N. Fuchs, G. Montambaux, and F. Pi\'echon}

\affiliation{
Laboratoire de Physique des Solides, CNRS UMR 8502, Univ. Paris-Sud, F-91405 Orsay cedex, France}

\begin{abstract}

The relativistic Landau levels in the layered organic material \BEDT~
[BEDT-TTF=bis(ethylenedithio)tetrathiafulvalene] 
are sensitive to the 
tilt of the Dirac cones, which, as in the case of graphene, determine the low-energy electronic properties 
under appropriate pressure. We show that an applied inplane electric field, which happens to be in 
competition with the tilt of the cones, lifts the twofold valley degeneracy due to a different level 
spacing. The scenario may be tested in infrared transmission spectroscopy.

\end{abstract}
\pacs{73.61.Wp, 73.61Ph, 73.43.-f}
\maketitle

The study of relativistic effects in condensed matter systems has become a major issue since the 
discovery of a particular quantum Hall effect in graphene \cite{novoselov,zhang}, which may be understood 
in terms of relativistic two-dimensional (2D) charge carriers at the Fermi level \cite{antonioRev}.
These relativistic carriers occur in the vicinity of the crystallographic points $K$ and $K'$ within the
first Brillouin zone (BZ), where the valence band touches the conduction band. The electronic energy
dispersion is then linear in the wave vector $k$, $\epsilon(k)=\hbar v_F k$, in terms of the Fermi velocity 
$v_F\sim 10^6$ m/s, which is roughly 300 times less than the velocity of light. 

Another system, less known than graphene, where 2D relativistic charge carriers are expected to describe 
the low-energy electronic properties, is the organic layered compound \BEDT~[BEDT-TTF=bis(ethylenedithio)tetrathiafulvalene]
\cite{katayama,kobayashi,fukuyama}.
The electronic structure is, however, more involved than in graphene due to the presence of four sites
per unit cell
with highly anisotropic overlap integrals. An additional complication stems from the presence of strong electronic
correlations in these materials, which lead to insulating phases with modulated charge density at low temperatures 
\cite{OrganicsRev}.
Fermi-liquid-type physics may be induced when pressure is applied to this 3/4-filled system, in which case the
Fermi level resides at the contact points of the upper two bands, which form Dirac cones \cite{katayama,kobayashi}. 
Correlation effects, as well as different on-site energies, may then be taken into account within a 
mean-field theory and renormalize the band parameters \cite{katayama2}.

A particularly
interesting feature of Dirac cones in \BEDT, which occur in two distinct points within the first BZ, 
is their {\sl tilt} in the momentum-energy space.
This tilt may be understood as due to an interplay between a highly anisotropic nearest-neighbor hopping,
which drives the Dirac points away from points of high crystallographic symmetry, and a next-nearest-neighbor 
hopping which is of comparable size as the nearest-neighbor one \cite{GFMP}. In
a magnetic field, the spectrum is organized in relativistic Landau levels (LL) as in graphene, but
the tilt of the Dirac cones reduces the effective Fermi velocity and, thus, the LL spacing \cite{GFMP,morinari}.

In spite of this theoretical understanding and in contrast to graphene, strong experimental 
evidence for (tilted) Dirac cones in \BEDT~under pressure is yet lacking. An unusual $T^2$ dependence of the carrier density,
observed in transport measurements \cite{kajita}, 
may be interpreted as an indication of relativistic 
carriers \cite{antonioRev,katayama}. This temperature dependence is in competition with a $T^{-2}$
dependence of the mobility, such that the electric conductivity is roughly temperature-independent.
Very recently, magneto-transport measurements have revealed the existence of a zero-energy LL in \BEDT~\cite{tajima2},
in agreement with theoretical predictions \cite{GFMP}. 

Another means of reducing the LL spacing (in graphene), different from the above-mentioned 
tilt of the Dirac cones due to a particular combination of the hopping parameters, 
has been proposed by Lukose, Baskaran, and Shankar \cite{lukose}. The application of an inplane electric field 
reduces indeed the LL spacing by a factor $[1-(E/v_F B)^2]^{3/4}$. This effect is a direct consequence 
of the covariance of the Dirac equation for fermions in an electromagnetic field and may be understood in terms of 
a Lorentz transformation into a reference frame, where the electric field vanishes.

Here, we show that the combination of an inplane electric field and a perpendicular magnetic field $\bB=B\be_z$
may be viewed as an effective change of the tilt of the Dirac cones. Indeed, 
it renormalizes the tilt parameter, which becomes not only dependent on the electric field but also on the 
valley index $\xi=\pm 1$. The renormalized tilt parameter affects the effective cyclotron energy and the LL
spectrum. The latter depends explicitly on the valley index, and, thus, the inplane electric field 
lifts the twofold valley degeneracy for the LLs $n\neq 0$. The central LL $n=0$ remains at zero energy in both valleys
and maintains its twofold valley degeneracy. 
A quantum Hall measurement, which one may first think of to reveal the lifted valley degeneracy,
is a difficult task due to the lack of single-layer \BEDT.
An alternative and more realistic experimental test of the proposed scenario may be 
infrared transmission spectroscopy, which has been proven to be a powerful tool in the determination of
relativistic LLs in graphene \cite{sadowski,spectroGraph2}.

In the absence of a magnetic field, the low-energy electronic properties of \BEDT~with weak layer coupling
may be described
by the minimal form of the generalized 2D Weyl Hamiltonian for the valley $\xi$ \cite{GFMP}
\beq\label{eq01}
H_{\xi}=\xi\left(\bw_0\cdot\bq\, \sigma^0+w_xq_x\sigma^x+w_yq_y\sigma^y\right),
\eeq
in terms of the $2\times 2$ Pauli matrices $\sigma^x$, $\sigma^y$, and $\sigma^0\equiv 1$
(we use a system of units with $\hbar=1$). The velocities 
$\bw_0=(w_{0x},w_{0y})$ and $\bw=(w_x,w_y)$ may be viewed as effective parameters. Estimates, calculated within
the model proposed by Hotta \cite{hotta} and based on the overlap integrals of Ref. \cite{mori}, yield 
$w_x\simeq 2.14$ eV\AA, $w_y\simeq 0.22$ eV\AA, $w_{0x} \simeq -0.0075$ eV\AA, and $w_{0y}\simeq 0.074$ eV\AA~\cite{GFMP}. 

In order to account for the perpendicular magnetic field $B\be_{z}=\nabla\times\bA$, one may use the Peierls substitution 
\beq\label{PeierlsSub}
\bq\rightarrow \Pib=\bq+e\bA,
\eeq
which is a valid approximation when the lattice spacing [$\sim 1$ nm in \BEDT] is much smaller than the magnetic
length $l_B=1/\sqrt{eB}\simeq 26/\sqrt{B{\rm [T]}}$ nm.
With the help of the ladder operators
\beqn\label{ladderOp}
\nn
a &=& \frac{l_B}{\sqrt{2w_xw_y}}\left(w_x\Pi_x-iw_y\Pi_y\right),\\
a^{\dagger} &=& \frac{l_B}{\sqrt{2w_xw_y}}\left(w_x\Pi_x+iw_y\Pi_y\right),
\eeqn
one obtains the Hamiltonian \cite{GFMP}
\beq\label{WeylB}
H_{\xi}=\xi\frac{\sqrt{ 2 w_xw_y}}{l_B}\left(\begin{array}{cc}
 \frac{\wtilde_0}{2}(a e^{i \varphi}+a^{\dagger}e^{-i \varphi} ) & a \\
 a^{\dagger} & \frac{\wtilde_0}{2}(a e^{i \varphi}+a^{\dagger}e^{-i \varphi} )
\end{array}
\right).
\eeq
where
$$\wtilde_0 e^{i\varphi}\equiv \frac{w_{0x}}{w_x}+i\frac{w_{0y}}{w_y} ,$$
in terms of the effective tilt parameter 
\beq\label{tiltparam}
\wtilde_0\equiv\sqrt{\left(\frac{w_{0x}}{w_x}\right)^2+
\left(\frac{w_{0y}}{w_y}\right)^2}. 
\eeq
The angle $\varphi$ denotes the angle between the $x$-axis and the direction of the effective tilt $(w_{0x}/w_x,w_{0y}/w_y)$, 
renormalized by the Fermi velocities $w_x$ and $w_y$ in the $x$- and $y$-direction, respectively.
A semiclassical \cite{GFMP} as well as a full quantum treatment \cite{morinari} of the Hamiltonian (\ref{WeylB})
yields the energy spectrum
\beq\label{eq02}
\epsilon_{\lambda, n}=\lambda \frac{\sqrt{w_xw_y}}{l_B}(1-\wtilde_0^2)^{3/4}\sqrt{2n}\, ,
\eeq
where $\lambda=\pm$ indicates the band index. The spectrum, which has the same structure as that of 
relativistic electrons in graphene, does not depend on the valley index $\xi$ and is, thus, 
twofold (fourfold, when one takes into account the electron spin) degenerate, in addition to the 
macroscopic LL degeneracy characterized by the density of flux quanta $n_B=1/2\pi l_B^2$.

In the presence of an inplane electric field $\bE=E\be_{\parallel}$ in the  
direction $\be_{\parallel}=(\cos\theta,\sin\theta)$ in the $xy$-plane, one needs to add a term 
$e\bE\cdot\br\, \sigma^0 = eEx'\, \sigma^0$ 
to the Hamiltonian (\ref{WeylB}), where $x'\equiv x\cos\theta + y\sin\theta$. 
For non-tilted Dirac cones ($\wtilde_0=0$) and for an
isotropic Fermi velocity $v_F=w_x=w_y$, as in the case of graphene, the electric field leads 
to the same energy spectrum (\ref{eq02}). One simply needs to replace the tilt parameter 
$\wtilde_0\rightarrow E/v_F B$ \cite{lukose} and 
to add a term $(E/B) k$, where $k$ is the wave vector in the direction 
$\be_{\perp}=(-\sin\theta,\cos\theta)$ perpendicular to $\be_{\parallel}$.
Here, one naturally chooses the Landau gauge $\bA=Bx'\be_{\perp}$,
such that the full Hamiltonian remains translationally invariant in
the $y'$-direction ($\be_{\perp}$) and the associated wave vector $k$ is, therefore, a good 
quantum number.

An inplane electric field in \BEDT, i.e. when the Fermi velocity is anisotropic
and the Dirac cones are tilted, is slightly more involved. The complication stems from the competition 
of three characteristic directions: the main axes of the anisotropic Fermi velocities $w_x$ and
$w_y$, which define the reference frame; the direction of the tilt described by $\varphi$;
and the direction of the applied electric field given by the angle $\theta$. One may then express the variable $x'=kl_B^2+\Pi_y'(a,a^{\dagger})l_B^2$, where
$\Pi_y'(a,a^{\dagger})=[-\sqrt{w_y/w_x}(a+a^{\dagger})\sin\theta+i\sqrt{w_x/w_y}(a-a^{\dagger})\cos\theta]$ is the 
$y'$-component of the gauge-invariant momentum (\ref{PeierlsSub}). The final
Hamiltonian in the presence of an electric field reads
\beq\label{WeylE}
H_{\xi}^{E/B}=H_{\xi}'+\frac{E}{B} k\,\sigma^0,
\eeq
where $H_{\xi}'$ is the same as that of Eq. (\ref{WeylB}) if one replaces the tilt parameter 
$\wtilde_0\exp(i\varphi)$ by 
\beq\label{eq03}
\wtilde_{\xi}(E) e^{i\varphi_{\xi}(E)}\equiv \frac{w_{\xi x}}{w_x}+i\frac{w_{\xi y}}{w_y} .
\eeq
Here, the renormalized tilt velocity is given by
\beq\label{eq04}
\bw_{\xi}=(w_{\xi x},w_{\xi y})\equiv \bw_0 - \xi \frac{\bE\times\bB}{B^2},
\eeq
and the angle $\varphi_{\xi}$ is related to the angle between the renormalized tilt velocity and the 
$x$-axis.

Eq. (\ref{eq04}) is the central result of this paper. The additional term in Eq. (\ref{eq04}) is nothing
other than the drift velocity $\bv_D$ of a charged particle in the presence of both a magnetic and an electric 
field \cite{jackson}. This drift velocity may therefore, on its own, tilt the original Dirac cones in the same
manner as the 
velocity $\bw_0$, which is due to the special combination of the nearest-neighbor and next-nearest-neighbor
overlap integrals in the tight-binding model. The intervening electric field depends effectively 
on the valley index $\xi$, and so does the LL spectrum,
\beq\label{eq05}
\epsilon_{\lambda, n;k}^{\xi}(E)=\lambda \frac{\sqrt{w_xw_y}}{l_B}\left[1-\wtilde_{\xi}(E)^2\right]^{3/4}\sqrt{2n} 
+ \frac{E}{B} k\, .
\eeq

One may distinguish three different instances for the LL spectrum. 

{\sl (i)} In the case with no original tilt ($\bw_0=0$), 
such as for unconstrained graphene, one recovers the valley-independent result of Ref. \cite{lukose}, where the 
decrease of the effective Fermi velocity is due to the Lorentz invariance of the system. Indeed, the 
effective magnetic field is reduced by a factor 
\beq\label{eqRapid}
\sqrt{1-(E/v_F B)^2}
\eeq
as a consequence of a Lorentz transformation into the
reference frame, which moves at the drift velocity and where the electric field vanishes. Because the
energy is proportional to $1/l_B\propto \sqrt{B}$, this leads to a reduction by a factor $[1-(E/v_F B)^2]^{1/4}$ 
of the LL spacing. However,
the spectrum needs to be multiplied by another factor (\ref{eqRapid}) when measured in the lab frame, due to the Lorentz transformation 
of the energy. The total reduction is, therefore, given by the
factor $[1-(E/v_F B)^2]^{3/4}$, in agreement with Eqs. (\ref{eq02}) and (\ref{eq05}). The LL spectrum collapses once the
drift velocity $v_D=E/B$ exceeds the Fermi velocity $v_F$, which is an upper bound in analogy with the velocity of light. 
Physically, the system enters then into a regime which is dominated by the electric field, and where a Lorentz transformation
may be found into a reference frame with no {\sl magnetic} field \cite{jackson}.

{\sl (ii)} In the case of originally tilted Dirac cones ($\bw_0\neq 0$) and no electric field, one recovers the LL spectrum
discussed in Refs. \cite{GFMP} and \cite{morinari}. As for the above case {\sl (i)}, the spectrum is valley-degenerate and collapses
beyond a critical tilt value given in terms of the tilt parameter $\wtilde_0=1$. Indeed, for $\wtilde_0<1$ the semiclassical
trajectories are closed orbits (ellipses), whereas those for $\wtilde_0>1$ are open parabolas, which would give rise to a 
continuous spectrum. The two regimes of closed and open orbits
may therefore be identified with the relativistic ones of the case {\sl (i)} dominated by the magnetic 
and the electric field, respectively.

\begin{figure}
\epsfysize+5.5cm
\epsffile{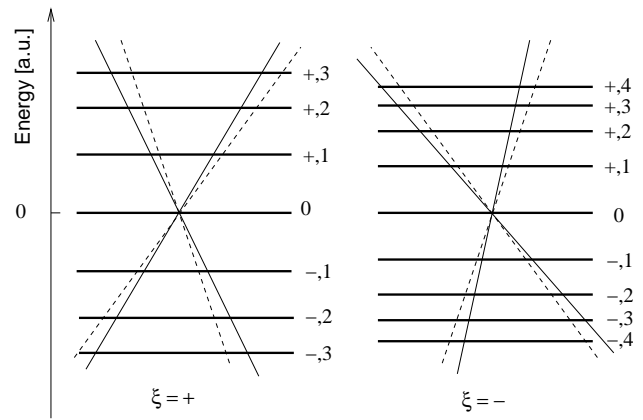}
\caption{Sketch of the valley-dependent LL spectrum for tilted Dirac cones in the presence of an electric field (thick lines). 
We omitted the inclination of the LLs due to the term $ (E/B)k$, which lifts the LL degeneracy.
The dashed lines schematically represent the tilted cones in the two valleys $\xi=+$ and $-$ in the absence of an 
electric field. The cones in the two valleys are tilted in opposite directions in the momentum-energy space, whereas
the electric field acts in the same direction. 
The LL spectrum in the presence of an electric field in $\xi=+$ ($\xi=-$) is that of a cone with a 
decreased (increased) tilt (full lines).}
\label{fig01}
\end{figure}

{\sl (iii)} The most interesting situation arises when both the electric field and tilted Dirac cones are present, 
such as in \BEDT. In this case,
the valley degeneracy is lifted, and the energy spectra are different for the two valleys. The effect is most pronounced when
$\bw_0 \perp \bE $, in which case the electric field redresses the cone in one valley 
(say $\xi=+$ \footnote{The role of the valleys is
exchanged if one inverts the direction of the electric field.}) while enhancing the tilt in the other one, $\xi=-$, (Fig. \ref{fig01}). This leads to a larger LL spacing in
$\xi=+$, whereas that in $\xi=-$ decreases, and the critical electric field, beyond which the LL spectrum collapses and 
becomes continuous, is different in the two different valleys. Notice that the zero-energy LL remains unaffected
by the electric field and maintains, therefore, its original twofold valley degeneracy. 

A particularly interesting value of the
electric field is $E_{comp}=|\bw_0| B$, when the drift velocity exactly compensates $\bw_0$. Notice that because 
$\bw_0$ is almost parallel to the $y$-axis ($\varphi\simeq 91^{\circ}$), the electric field is then to be chosen in the
$x$-direction. In this case, the LL spacing is
maximal in one valley, with a cyclotron energy of $\epsilon_{max}=\sqrt{2w_xw_y}/l_B$, whereas it is 
$\epsilon_{min}=\sqrt{2w_xw_y}[1-4\wtilde_0^2]^{3/4}/l_B$ in the other one. With the above-mentioned values of the velocities, one has
$|\bw_0|\sim 10^5$ m/s, and the compensation condition is, therefore, fulfilled for an electric field of 
$E_{comp}\sim 10^6$ V/m at a typical magnetic field of $10$ T. Inplane electric fields of this strength (and larger) 
have indeed been obtained in graphene \cite{bachtold}, and one may expect that they
are in the accessible experimental range also in \BEDT, which is expected to have similar screening properties
as graphene. A rough
estimate of the tilt parameter in \BEDT~yields $\wtilde_0\sim 0.33$ \cite{GFMP}, and one may therefore expect a rather large 
ratio $\epsilon_{min}/\epsilon_{max}\sim 0.6$, i.e. a 40\% modulation.

As an experimental test of the proposed scenario, one may first think of a quantum Hall measurement, where one would
expect that the plateau series at the filling factors
$\nu=n_{el}/n_B=\pm 2(2n+1)$ of the relativistic quantum Hall effect evolves into a series of $\nu=\pm 2(n+1)$ once a sufficiently
large electric field is applied. Here, $n_{el}$ is the 2D carrier density,
and we have taken into account
the twofold spin degeneracy. Such a measurement would, however, require a single layer of \BEDT, which has not been 
fabricated yet.
Another more realistic experimental test of the proposed scenario consists of an infrared spectroscopic measurement, which has
been performed with great succes in epitaxial \cite{sadowski} and exfoliated graphene \cite{spectroGraph2} and which 
reveals the relativistic LL structure of these materials. The dipolar coupling is sensitive to the transitions 
$\lambda,n\rightarrow \lambda',n\pm 1$, and one may therefore access both intraband ($\lambda'=\lambda$) and interband
($\lambda'=-\lambda$) transitions. In the case of 3/4-filled \BEDT, where the $n=0$ LL is half-filled,
one would expect absorption lines at energies (for all integers $n$)
\beq\label{eq10}
\Delta_n^\xi (E) = \frac{\sqrt{2 w_xw_y}}{l_B}\left[1-\wtilde_{\xi}(E)^2\right]^{3/4}(\sqrt{n}+\sqrt{n+1}),
\eeq
which split into two lines each when an inplane electric field is applied. A linear expansion in $E$ of Eq. (\ref{eq10})
yields a relative splitting $\eta=[\Delta_n^+(E)-\Delta_n^-(E)]/\Delta_n^{\xi}(E=0)\sim E/v_F^*B$ 
of the lines if the direction of $\bE$ is appropriately chosen (close to the
$\be_x$ direction, where the tilt is maximal) and if the LL broadening is sufficiently small. Here, $v_F^*=\sqrt{w_xw_y}[1-\tilde{w}_{\xi}(E)^2]^{3/4}\simeq 10^5$ m/s is the renormalized Fermi velocity which accounts for
the tilt as well as for the anisotropy of the Dirac cones in $\alpha$-(BEDT-TTF)$_2$I$_3$.

Notice, however, that the strength of the inplane electric field is delimited by breakdown processes, which
may rearrange the LL filling \cite{nachtwei}. Above a critical electric field $E_c$, once the associated potential energy 
acquired over the characteristic length $l_B$ is a substantial fraction $\gamma$ of the inter-LL spacing $\delta_n
=\epsilon_{n+1}-\epsilon_n$, 
$eE_cl_B=\gamma \delta_n$, an electron may tunnel from one LL $n$ to a neighboring one $n+1$.
In GaAs heterostructures, $\gamma\simeq 0.1$ is a reasonable estimate 
to account for the observed breakdown \cite{tsemekhman}. If one considers a similar limitation in \BEDT, the 
breakdown occurs at $E_c\sim \gamma v_F^* B \sim 10^{5}\ {\rm V/m}\ < E_{comp}$, and the maximal relative strength of the
valley-degeneracy lifting may be estimated from (\ref{eq10}) as $\gamma$, which yields a $\sim 10\%$ 
effect, for a cyclotron energy of roughly $\epsilon\sim 3.4\sqrt{B{\rm [T]}}$ meV 
(10 meV at $B=10$ T).

We finally point out that a similar valley-degeneracy lifting may occur un graphene, where a tilt of the Dirac cones
may, in principle, be induced under uniaxial strain \cite{GFMP}. The effective tilt parameter has beeen estimated as 
$\wtilde_0\sim 0.6\varepsilon$, where $\varepsilon$ is the relative variation of the bond length under uniaxial strain \cite{GFMP}, which
may be on the order of 10-20\% before the graphene sheet cracks \cite{strain}. The order of magnitude for lifting the valley 
degeneracy is $\eta \sim(E/B v_F)\varepsilon$, and if one assumes a breakdown parameter $\gamma\sim 0.1$, this yields
a $\sim 1\%$ effect for a critical electric field of roughly $10^6$ V/m, which is roughly ten times larger than 
that of \BEDT, due to the larger Fermi velocity in graphene.


In conclusion, we have shown that a simultaneous tilt of the Dirac cones and an applied electric field in the 2D plane
is capable of lifting the twofold valley degeneracy of the LL spectrum. Whereas one may induce a tilt of 
the Dirac cones in graphene by application of a uniaxial strain \cite{GFMP}, 
those in \BEDT~are naturally tilted under moderate pressure, due to a particular 
interplay between the nearest-neighbor and next-nearest-neighbor hopping integrals \cite{katayama,kobayashi,fukuyama,GFMP}.
The inplane electric field acts, in the presence of a perpendicular magnetic field, as an additional effective tilt, 
which is valley-dependent and which competes with the original tilt. If the direction of the inplane field is properly
chosen, it redresses the Dirac cone in one valley $\xi$ while enhancing the tilt in the other one $-\xi$. 
The LL spectrum is therefore stretched in $\xi$ and compressed in $-\xi$. The effect may be observable in 
infrared transmission spectroscopy.

This work is partially supported by the Agence Nationale de la Recherche under Grant No. ANR-06-NANO-019-03.

\end{document}